# Evaluating the Performance of Some Local Optimizers for Variational Quantum Classifiers

**Nisheeth Joshi[1], Pragya Katyayan[1], Syed Afroz Ahmed[2]**

[1]Department of Computer Science, Banasthali Vidyapith, India
[2]Umm Al Quwain University, United Arab Emirates

Email: nisheeth.joshi@rediffmail.com

**Abstract**. In this paper, we have studied the performance and role of local optimizers in quantum variational circuits. We studied the performance of the two most popular optimizers and compared their results with some popular classical machine learning algorithms. The classical algorithms we used in our study are support vector machine (SVM), gradient boosting (GB) and random forest (RF). These were compared with a variational quantum classifier (VQC) using two sets of local optimizers viz AQGD and COBYLA. For experimenting with VQC, IBM Quantum Experience and IBM Qiskit was used while for classical machine learning models, scikit-learn was used. The results show that machine learning on noisy immediate scale quantum machines can produce comparable results as on classical machines. For our experiments, we have used a popular restaurant sentiment analysis dataset. The extracted features from this dataset and then after applying PCA reduced the feature set into 5 features. Quantum ML models were trained using 100 epochs and 150 epochs on using EfficientSU2 variational circuit. Overall, four Quantum ML models were trained and three Classical ML models were trained. The performance of the trained models was evaluated using standard evaluation measures viz, Accuracy, Precision, Recall, F-Score. In all the cases AQGD optimizer-based model with 100 Epochs performed better than all other models. It produced an accuracy of 77% and an F-Score of 0.785 which were highest across all the trained models.

## 1. Introduction

Our understanding of machine learning (ML) (and deep learning (DL) concepts has come of age. Mostly everything that can be studied has already been studied and now we are moving towards optimizing the performance of the ML/DL models. This quest for continuous advancement has made us look into quantum mechanics where we try to couple the concepts of quantum physics into machine learning. The synergy of these two concepts had given rise to a new area of study is termed quantum machine learning (QML). Until now ML/DL algorithms could process a high volume of data and based on it could make intelligent decisions. QML has made the process even better as now we can learn new patterns from data that were previously termed as computational impossible. This is achieved by converting the classical input data into quantum states, thus making the process easier for a quantum computer to extract patterns of interest from them.

In quantum settings, algorithms can be of two types. There are classical algorithms and quantum algorithms. Classical algorithms are ones that use mostly classical constructs as used in classical machines. Quantum algorithms work with quantum states to make predictions. Similarly, data can be of two types. It can be classical data or quantum data. Thus, four different types of machine learning models can be developed using this information. These are ML models with classical algorithms working on classical data. ML modes with classical algorithms with quantum data. ML models with quantum algorithms with classical data and ML models with quantum algorithms with quantum data. Figure 1 shows these four types of models with two sets of algorithms (classical and quantum) and two types of data (classical and quantum).

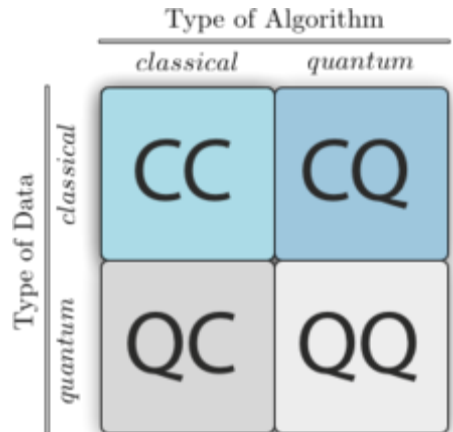

Figure 1: Four different approaches to combine the disciplines of quantum computing and machine learning [1]

The rest of the paper is organized as: section 2 shows the work done in the area of quantum machine learning. Section 3 shows our experiments, the design of a quantum classification algorithm and how we tuned it specifically to our dataset. Section 4 shows the results of classification from quantum machine learning algorithms and classical machine learning algorithms and how good or bad did the quantum algorithm performed in comparison to its classical counterparts. Finally, Section 5 concludes the paper.

## 2. Literature Review

Since we are quantum computers are still under development, so is quantum machine learning. At the current state of the art, we can perform are experiments either on quantum simulators or noisy immediate scale quantum (NISQ) computers which are still being researched. We still have a few years where we can have a fully functional noise-tolerant quantum computer, but this should not stop us from the joy of exploring new possibilities. As quantum computing is advancing each day so does quantum machine learning.

Kak [2] showed how a quantum neural network can be used and how it would more versatile than classical artificial neural networks. Menneer and Narayanan [3] showed how multiple single-layered neural networks can be used to form a much larger quantum neural network. Perus [4] was the first to show the use of gradient descent with quantum gates (more specifically CNOT gates) and developed a mechanism to facilitate parallelism in quantum neural networks. Faber et al. [5] argued how artificial neural networks can be implemented on quantum hardware. Schuld et al. [6] showed how quantum neural networks can be used in classification by using a distance function. By doing so they showed how quantum phenomenon can be in line with quantum theory.

Weber at el. [7] showed how quantum machine learning (specifically quantum neural networks) can make use of classical computers to model classification problems. Mitarai et al. [8] proposed a hybrid algorithm for quantum machine learning which they termed quantum circuit learning and showed how it can be tuned by using parameters. They further showed how a high-depth circuit can be implemented.

They theoretically showed the performance of their circuit learning and also did mathematical simulations to prove this. Abhijit et al. [9] showed how programmers working on classical computers can be trained to program quantum computers. They developed a complete tutorial for this. Kopczyk [10] showed how data science can be used in quantum computing in general and quantum machine learning in particular for their analysis tasks. They made a detailed step by step explanation of all the algorithms that are being used in quantum machine learning.

Schuld et al. [11] did an evaluation of gradients on quantum computers. They showed how the variational algorithm performs when gradients are used for optimization. They showed procedure of estimating expectation values of gradients for quantum measurements. Fastovets et al. [12] proposed approaches through which classical machine learning algorithms can be executed on quantum computers. They demonstrated their approach by executing a multiclass tensor network algorithm on a quantum computer provided by IBM Quantum Experience. Further, they showed how quantum tomography problem can be approached through artificial neural networks. They showed how their approach can predict the quantum state by reducing the noise. Zhao et al. [13] showed how the problem of large measurement computation in variational algorithms can be reduced using constant factor speedup for lattice and random Pauli Hamiltonians and showed promising results.

Lu at el. [14] adversarial machine learning on quantum computers. They showed how quantum machine learning-based classifiers which produce good results can be deceived by adversarial examples that use imperceptible perturbations to the original legitimate samples. Lloyd et al. [15] proposed quantum metric learning where state preparation tasks in a quantum circuit learning can be minimized and would also help in better measurement calculation. Terashi et al. [16] did a comparison of classical machine learning algorithms with quantum machine learning algorithms in high-energy physics applications. Through experiments, they showed that quantum machine learning can be used for this task. They specifically studied variational quantum algorithms.

Wu et al. [17] showed how a robust machine learning algorithm can be used on a noisy immediate scale quantum computer. In their algorithm, they incorporate feature selection by encoding input data to quantum states. Zhang et al. [18] argued that quantum machine learning algorithms which show speed up over classical machine learning algorithm could not keep this advantage while decoding the quantum states into final classical output. They proposed a protocol to address this issue by using the Gram-Schmidt orthonormal procedure. Garg and Ramakrishnan [19] reviewed advances made in quantum deep learning. They showed how quantum deep neural networks have advanced and also showed how can be used in natural language processing tasks. Guan et al. [20] investigated if quantum machine learning approaches can be used in the area of high-energy physics. They gave a very detailed description of this by providing a review of the work done in this area. Guan et al. [21] developed an algorithm that could verify if a quantum machine learning algorithm is robust on a particular set of training data. Through experiments, they showed how this approach improves the accuracy of the quantum machine learning model.

Suzuki and Katouda [22] showed how parameterized quantum circuits can be used to predict toxicity in organic chemistry. They studied the toxicity of some phenol structures. Through experiments, they showed that their quantum machine learning approach performed significantly better than the existing linear regression-based methods. Blance and Spannowsky [23] studied the use of quantum machine learning in particle physics. They proposed an algorithm using a quantum variational circuit that used multiple gradient descent approaches which optimized their results. They applied their algorithm on resonance search in di-top final states and showed that their approach performs better than the approach is being used currently.

This is becoming promising as large corporations like Google [24] and IBM [25] are researching quantum computers and have also provided open-source libraries to play with their quantum hardware. More specifically with quantum machine learning and test the limits of their hardware. Recently Google has also released Tensorflow Quantum which is built on top of their popular Tensorflow libraries and can be used to play with quantum machine learning.

## 3. Methods and Materials

This section explains the experiment that was performed using Qiskit Toolkit which is developed for experimenting with noisy immediate scale quantum (NISQ) machines. We have applied sentiment analysis for our experimentation and compared the performance of quantum machine learning models with classical machine learning models. We have checked the performance of two local optimizers of quantum machine learning and have tried to identify which one gives better performance on our dataset.

*3.1. Pre-processing*

For training any machine learning model, the first step is pre-processing as it helps in the extraction and selection of features. We have used a restaurant review dataset which has positive and negative reviews. This corpus had 500 positive and 500 negative reviews. Since the data was textual, we extract features from these textual features. The first feature that we extracted was the number of punctuations that the dataset had. Next, we calculated the length of the review. Then we removed the stop words from the corpus and calculated the tf-idf frequencies for unigram lexicons in the corpus. This generated 1698 tf-idf frequencies (features). This was a matrix of 1000x1698 data points, which had 1000 rows and 1698 columns.

For training a machine learning model on a computer, we need to assign one feature to one qubit. Since the current state of the art quantum machines does not provide such a high number of qubits, we were needed to scale down the number of features so that they can be processed with quantum computers (Maximum features that can be supported are thirty-two). Thus we transformed this 1000x1698 data point matrix into a matrix of 1000x1 by multiplying a unitary matrix of 1000x1.

The following explanation shows the working of this process, where we have a matrix of m rows and n columns which needs to be transformed into a matrix of m rows and 1 column.

$$\begin{bmatrix} a_{11} a_{12} a_{13} \cdots & \cdots & a_{1n} \\ a_{21} a_{22} a_{23} \cdots & \cdots & a_{2n} \\ \vdots & \vdots & \vdots \\ \vdots & \vdots & \vdots \\ a_{m1} a_{m2} a_{m3} \cdots & \cdots & a_{mn} \end{bmatrix}_{m \times n} \times \begin{bmatrix} 1 \\ 1 \\ \vdots \\ \vdots \\ 1 \end{bmatrix}_{n \times 1}$$

$$= \begin{bmatrix} a_{11} \times 1 + a_{12} \times 1 + a_{13} \times 1 + \cdots & \cdots & + a_{1n} \times 1 \\ a_{21} \times 1 + a_{22} \times 1 + a_{23} \times 1 + \cdots & \cdots & + a_{2n} \times 1 \\ \vdots & \vdots & \vdots \\ \vdots & \vdots & \vdots \\ a_{m1} \times 1 + a_{m2} \times 1 + a_{m3} \times 1 + \cdots & \cdots & + a_{mn} \times 1 \end{bmatrix}_{m \times 1}$$

$$= \begin{bmatrix} a_{11} + a_{12} + a_{13} + \cdots & \cdots & + a_{1n} \\ a_{21} + a_{22} + a_{23} + \cdots & \cdots & + a_{2n} \\ \vdots & \vdots & \vdots \\ \vdots & \vdots & \vdots \\ a_{11} + a_{12} + a_{13} + \cdots & \cdots & + a_{1n} \end{bmatrix}_{m \times 1} \quad (1)$$

Similarly, we computed the tf-idf frequencies for bigram and trigram word sequences. Other features that we extracted from the dataset were:

1. Number of Stopwords.
2. Number of Content Words
3. Number of Nouns
4. Number of Verbs

5. Number of words on which lemmatization was done
6. Number of words with all letters capital
7. Number of elongated words
8. Number of positive emoticons
9. Number of negative emoticons
10. Number of neutral emoticons
11. Number of slang words
12. Number of negation words
13. Total positive score of words in the negated context
14. Total negative score of words in negated context
15. Total positive score of words in a positive context
16. Total negative score of words in a negative context

Finally, we had a total of 17 features in our feature set. Next, we performed feature selection using principal component analysis (PCA), which reduced the features to just five features. With these five features, we had trained our classical and quantum classifiers.

*3.2. Structure of a Variational Quantum Classifier*

Variational Quantum Classifiers (VQC) are artificial neural networks. They are very popular as one can easily train a classifier without performing error correction which is required while working on NISQ machines as these machines tend to add noise in the output. VQC is considered as a hybrid classifier where a part of the processing is done on a classical computer viz parameter optimization and updation, and the cost calculation is done at the quantum computer which finally helps in calculating the error and accuracy of the model.

Thus, to perform machine learning using VQC, we need to design a quantum circuit that can act as a machine learning classifier. This is achieved by designing a circuit which can have multiple parameters that are optimized to produce a minimum loss. The trained circuit can be shown using equation 2.

$$f(w, b, x) = y \quad (2)$$

Here, $f$ is the classifier that we are training through VQC and $y$ is the output label that we which to produce through our classifier. The objective of the function (classifier) $f$ is to reduce the loss function. This is similar to the artificial neural networks that we train in classical machine learning. The classifier is a circuit-centric quantum classifier [26] which has three sub-parts viz (1) state preparation circuit, (2) model circuit and (3) measurement and post-processing.

The initial state or state preparation circuit takes the feature vector as an input with $n$ features and encodes them into $n$ qubits. The model circuit is the main unit which applies various quantum gates to this input state and tires to minimize the loss. Finally, we measure the output label in the third sub-circuit which performs measurement and post-processing on the output received. The entire working of this entire VQC is shown in figure 2.

In figure 2, the feature map and variational circuit are the elaborations of the model circuit of figure 1. In both the figures, our dataset is transformed into a feature vector $\vec{f}$ which is then supplied for state preparation circuit which then converts the features in feature vector into qubits. These qubits are then sent to the model circuit where we qubits are first sent the quantum feature map which is a black box. The role of this box is to encode classical data $x_i$ into quantum state $|\varphi(x_i)>$. This is done by transforming the ground state $|0>^n$ into products of single and unitary phase gates. Here,

$$V_\emptyset(x) = V_\emptyset(x) H^{\otimes n} \quad (3)$$

Here H represents a Hadamard gate. The final computation is done by:

$$V_\emptyset(x) = exp\left(i \sum_{S \subseteq |n|} \emptyset s(\vec{x}) \prod_{i \in S} Z_i\right) \quad (4)$$

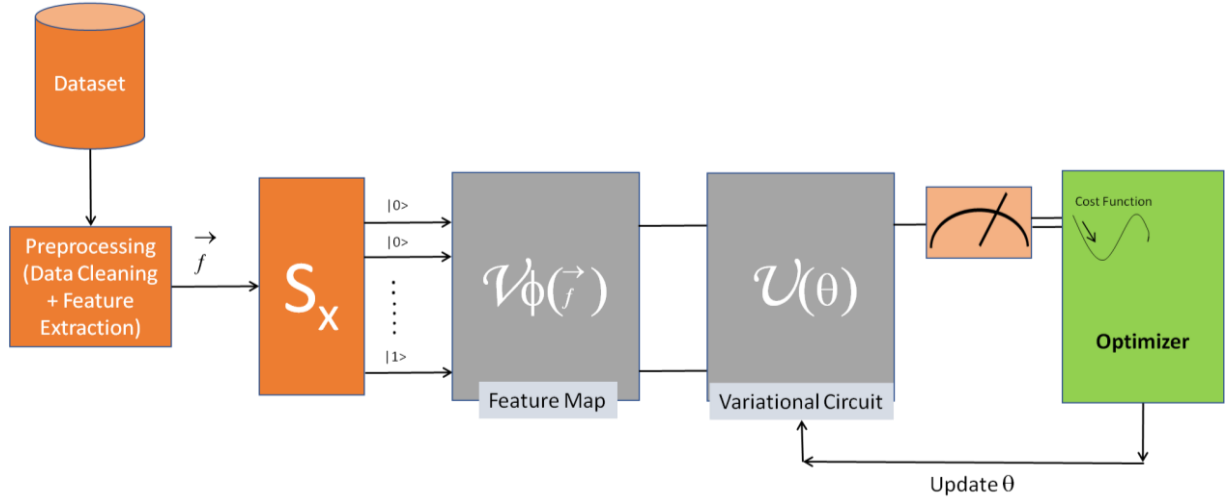

Figure 2: Block Diagram of a Variational Quantum Classifier

Here $V_\emptyset(x)$ is a diagonal gate which assumes Pauli-Z. Next, the results of this process are supplied to the variational circuit U(θ) which had $l$ layers of θ-parameters which are optimized during training by minimizing the cost function in the classical machine and thus tuning θ recursively. This parameterized approach to machine learning in a quantum computer is also referred to as quantum circuit learning [26] (QCL). QCL uses multiple exponential functions with the n qubits from the parameterized circuit. This is something that is not possible on classical machines; thus, this provides the quantum advantage over classical computing, as here we can represent a larger set of complex functions than what the classical computers can handle.

*3.3. Design of Variational Quantum Classifier for Sentiment Analysis using Custom (Parameterised) Circuits*

In order to develop a classifier, we were first required to design a circuit that could learn the patterns. In order to do so, we first configured some parameters which were required to train the final circuit. These were:

*3.3.1. Feature Map:* A feature map is the fundamental building block in a machine learning setup. It is a function that maps an input feature vector to feature space. We have been using feature maps in pattern recognition and computer vision-based systems. Feature maps are used because they help the learning algorithm to perform better and predict accurately. This is so because feature maps help in dimensionality reduction as it reduces the resources required in defining the enormous data. Moreover, if larger data is used, then it might cause overfitting of the model. To some extent, feature maps keep a check on this.

Broadly they have been used in several learning algorithms, but have gained popularity with the advent of kernel methods where they have been extensively used. With the dawn of deep learning systems, there has been a renewed interest in the machine learning community in customizing a variety of feature maps.

Since VQC is a kind of a neural network, it also uses a feature map. Many feature maps are available in qiskit, but we have used Pauli feature maps for our experiments. Equation 4 shown above is the formulation of a Pauli feature map. The $Z_i$ in this equation denotes a Pauli matrix. In our experiment, we have used this feature map and have customized it with 5 input strings (one each for 5 qubits). Figure 3 shows the circuit diagram for these feature maps. We have used 3 repetitions of this feature map in our experiment which means the circuit in figure 3 is drawn (used) three times in our classifier.

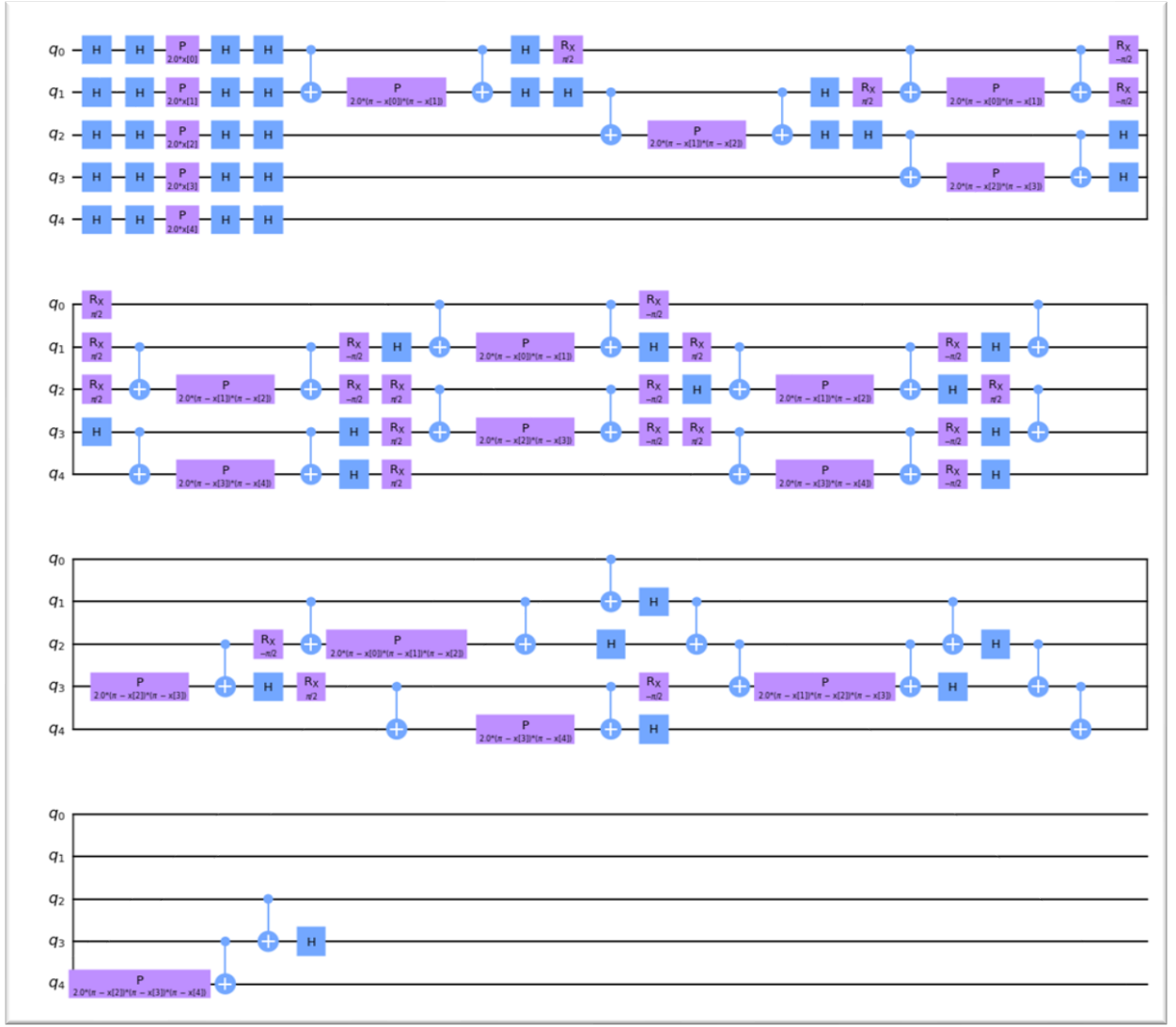

Figure 3: Circuit of Feature Map for Sentiment Analysis Task

*3.3.2. Variational Circuit*: A variational quantum circuit is a computational routine consisting of coherent quantum operations on quantum data, such as qubits, and concurrent real-time classical computation. It is an ordered sequence of quantum gates, measurements and resets, all of which may be conditioned on and use data from the real-time classical computation.

In our experiment, we have used the EfficientSU2 variational circuit. This circuit was used with a Pauli feature map with two different sets of epochs (100 and 150). Figure 4 shows the diagram for the EfficientSU2 variational circuit.

*3.3.3. Optimizer and Measurement*: In VQC the optimizer gets the results from the classifier and compares it with the actual result. Based on this it calculates the error (loss) in the training process and thus sends the feedback to the classifier which helps in tuning the weights (feature-map parameters) which can retain the variational circuit. This process is repeated until the loss is minimized. Finally, the optimizer results in the list of parameters that have the minimum error (loss) in training. In our experiment, we have used two different optimizers viz. COBYLA [27] and AQGD [28]. COBYLA (Constrained Optimization By Linear Approximation) optimizer is a numerical optimization method used to minimize the loss in constraint problems where the derivative of the objective is not known. Here the optimization is performed by linear approximation which is applied on constraints and the objective function. On the other hand, AQGD (Analytic Quantum Gradient Descent) optimizer is a quantum variant of the popular Gradient Descent function used in artificial neural networks.

Thus, we have trained four quantum classifiers, with each epoch of our variational circuit, we have used two optimizers i.e. Efficient SU2 with 100 Epochs and COBYLA optimizer, EfficientSU2 with 150 epochs and COBYLA optimizer, Efficient SU2 with 100 Epochs and AQGD optimizer, and finally, EfficientSU2 with 150 epochs and AQGD optimizer,

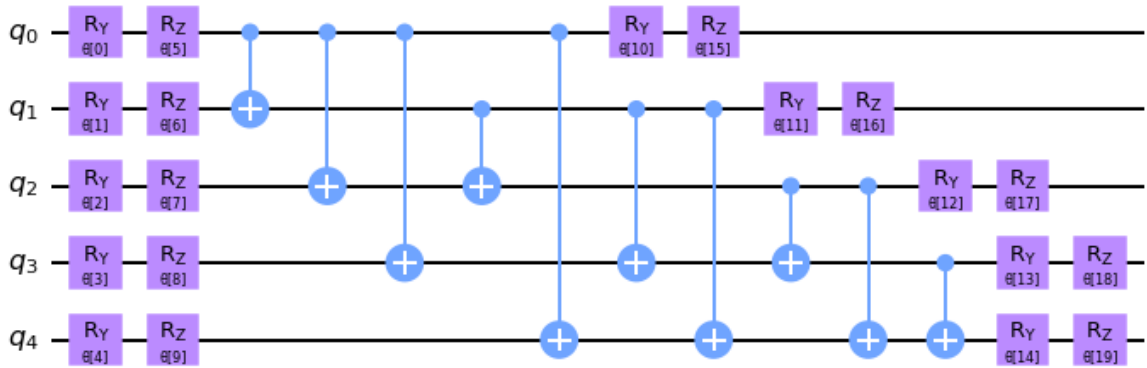

Figure 4: EfficientSU2 Variational Circuit for Sentiment Analysis Task

*3.4 Classical Machine Learning Algorithms:* In order to compare the results of our quantum machine learning classifier, we also trained three popular classical machine learning models viz support vector machines, random forest and gradient boosting. Among these gradient boosting and the random forest are considered as ensemble classifiers. Support vector machines on the other hand is a popular machine learning algorithm that has shown promising results in several machine learning tasks.

**4. Results and Discussion**

We had a corpus of 1000 data points with 500 being positive reviews and 500 being negative reviews. Among these, we divided the corpus into 3 sets viz training, validation and the test set. The training set and the validation set were used at the time of training the models. The models were trained on a training set and were optimized using the validation set. Once the models were trained, we used the test set which was completely unseen by the models. This gave fair chances to all the models and hence negated the possibility of biasness of data towards any particular model. The statistics of the three sets are shown in table 1.

| Set | Data points |
| --- | --- |
| Training Set | 600 |
| Validation Set | 200 |
| Test Set | 200 |

Table 1: Statistics of Training, Validation and Test Sets

Once the models were tested, we generated the confusion matrix of each of the models where we calculated the true positives, true negatives, false positives and false negatives. Table 2 shows a general confusion matrix where we can see how the results of the model are compared with the actual predictions. Here, TP is the number of true positives i.e. the positive tuples which were correctly identified by the model. TN is the number of true negatives i.e. the negative tuples which were correctly identified. FP is the number of false positives i.e. the negative tuples which were incorrectly identified as the positive tuples and FN is the number of false negatives i.e. the positive tuples which were incorrectly identified as negative. P is the total actual positive tuples and N is the total actual negative tuples. P' is the total positive tuples that the model has predicted and N' is the total negative tuples that the model has predicted.

|  |  | Predicted | | |
|---|---|---|---|---|
|  |  | Positive | Negative | Total |
| Actual | Positive | TP | FN | P |
| | Negative | FP | TN | N |
| | Total | P' | N' | P+N |

Table 2: General Confusion Matrix

|  |  | Predicted | | |
|---|---|---|---|---|
|  |  | Positive | Negative | Total |
| Actual | Positive | 64 | 45 | 109 |
| | Negative | 26 | 65 | 91 |
| | Total | 90 | 110 | 200 |

Table 3: Confusion Matrix of Gradient Boosting Model

|  |  | Predicted | | |
|---|---|---|---|---|
|  |  | Positive | Negative | Total |
| Actual | Positive | 67 | 42 | 109 |
| | Negative | 23 | 68 | 91 |
| | Total | 90 | 110 | 200 |

Table 4: Confusion Matrix of Random Forest Mode

The confusion matrix of Gradient Boosting is shown in table 3. Here we had 64 true positives, 45 false negatives, 26 false positives and 65 true negatives. The confusion matrix of Random Forest is shown in table 4. Here we had 67 true positives, 42 false negatives, 23 false positives and 68 true negatives. Table 5 shows the confusion matrix of the Support Vector Machines Model. Here we had 81 true positives, 28 false negatives, 29 false positives and 62 true negatives. Table 6 shows the confusion matrix of the EfficientSU2 Model with 100 epochs and COBYLA optimizer. Here we had 81 true positives, 28 false negatives, 23 false positives and 68 true negatives.

|  |  | Predicted | | |
|---|---|---|---|---|
|  |  | Positive | Negative | Total |
| Actual | Positive | 81 | 28 | 109 |
| | Negative | 29 | 62 | 91 |
| | Total | 110 | 90 | 200 |

Table 5: Confusion Matrix of Support Vector Machines Model

|  |  | Predicted | | |
|---|---|---|---|---|
|  |  | Positive | Negative | Total |
| Actual | Positive | 81 | 28 | 109 |
| | Negative | 23 | 68 | 91 |
| | Total | 104 | 96 | 200 |

Table 6: Confusion Matrix of EfficientSU2 Model with 100 Epochs and COBYLA Optimizer

Table 7 shows the confusion matrix of the EfficientSU2 Model with 150 epochs and COBYLA. Here we had 74 true positives, 35 false negatives, 21 false positives and 70 true negatives. Table 8 shows the confusion matrix of the EfficientSU2 Model with 100 epochs with AQGD Optimizer. Here we had 84 true positives, 25 false negatives, 21 false positives and 70 true negatives.

|        |          | Predicted |          |       |
|--------|----------|-----------|----------|-------|
|        |          | Positive  | Negative | Total |
| Actual | Positive | 74        | 35       | 109   |
| Actual | Negative | 21        | 70       | 91    |
|        | Total    | 95        | 105      | 200   |

Table 7: Confusion Matrix of EfficientSU2 Model with 150 Epochs and COBYLA Optimizer

|        |          | Predicted |          |       |
|--------|----------|-----------|----------|-------|
|        |          | Positive  | Negative | Total |
| Actual | Positive | 84        | 25       | 109   |
| Actual | Negative | 21        | 70       | 91    |
|        | Total    | 104       | 96       | 200   |

Table 8: Confusion Matrix of EfficientSU2 Model with 100 Epochs and AQGD Optimizer

Table 9 shows the confusion matrix of the EfficientSU2 Model with 150 epochs and AQGD Optimizer. Here we had 79 true positives, 30 false negatives, 23 false positives and 68 true negatives.

|        |          | Predicted |          |       |
|--------|----------|-----------|----------|-------|
|        |          | Positive  | Negative | Total |
| Actual | Positive | 79        | 30       | 109   |
| Actual | Negative | 23        | 68       | 91    |
|        | Total    | 101       | 99       | 200   |

Table 9: Confusion Matrix of EfficientSU2 Model with 150 Epochs

The results of the seven models are shown in table 10. We evaluated the performance of the models based on 5 evaluation parameters. These were accuracy, specificity, precision, recall and f-score. The accuracy of the model over a test set is tuples that are correctly classified by the model. It calculated using equation 5. Precision and Recall are the most popular measures which are used in the evaluation of classification models. Precision is the percentage of tuples labeled as positive, which are positive. It is also known as the measure of exactness and is calculated using equation 6. While Recall is the percentage of positive tuples that are labeled as such. It is also known as the measure of completeness and is calculated using equation 7. F-score is a combination of precision and recall. This combines them into a single measure so that we may calculate exactness and completeness at the same time. This is calculated using equation 8.

$$Accuracy = \frac{TP + TN}{P + N} \qquad (5)$$

$$Precision = \frac{TP}{TP+FP} \tag{6}$$

$$Recall = \frac{TP}{TP+FN} \tag{7}$$

$$F-Score = \frac{2 \times Precision \times Recall}{Precision + Recall} \tag{8}$$

|  | Accuracy | Precision | Recall | F-Score |
|---|---|---|---|---|
| Gradient Boosting | 0.645 | 0.7111 | 0.5872 | 0.6432 |
| Random Forest | 0.675 | 0.7444 | 0.6147 | 0.6734 |
| Support Vector Machine | 0.715 | 0.7364 | 0.7431 | 0.7397 |
| EfficientSU2 100 Epochs with COBYLA | 0.745 | 0.7788 | 0.7431 | 0.7605 |
| EfficientSU2 150 Epochs with COBYLA | 0.72 | 0.7789 | 0.6789 | 0.7255 |
| EfficientSU2 100 Epochs with AQGD | **0.77** | **0.8** | **0.7706** | **0.785** |
| EfficientSU2 150 Epochs with AQGD | 0.735 | 0.7745 | 0.7248 | 0.7488 |

Table 10: Evaluation results of Classical and Quantum Machine Learning Models

From the above table, we can see that EfficientSU2 with 100 epochs and AQGD Optimizer produced better results than the other models. The accuracy of this model was 77% which was the highest among all other models. This model also had a better precision, recall, and f-score with 80%, 77.06%, and 78.5% respectively.

## 5. Conclusion and Future Work

In this paper, we tried analyzing the performance of quantum machine learning models with classical machine learning models. We found that quantum ML models perform slightly better than classical machine learning models. We had used three classical machine learning models and compared their performance with four quantum machine learning models. We used four quantum machine learning models viz EfficientSU2 with 100 and 150 epochs with COBYLA and AQGD optimizers. Through experiments, we found that the AQGD optimizer with 100 epochs performed better than all the other models.

As an extension to this study, we would like to perform an experiment with other parameters and further improve the performance of our quantum machine learning model. We would like to experiment with different optimizers and feature maps and see the change in performance. Further, we would also like to experiment with datasets in different domains so that we may establish the true quantum advantage of QML models over classical ML models. When will have the actual noiseless quantum computers, we would also like to perform these experiments on them as well.

**Acknowledgment**
We would like to thank the IBM Quantum team for providing the access to the quantum computers through IBM Quantum Experience and also the Qiskit library which helped us in performing these experiments.